**Diagnostics of low and atmospheric pressure plasmas by means of mass spectrometry**

*J. Benedikt, D. Ellerweg, A. von Keudell*
Research Department Plasmas with Complex Interactions
Ruhr-University Bochum,
Germany

**Abstract**

The knowledge of absolute fluxes of reactive species such as radicals or energetic ions to the surface is crucial for understanding the growth or etching of thin films. These species have due to their high reactivity very low densities and their detection is therefore a challenging task. Mass spectrometry is a very sensitive technique and it will be demonstrated that it is a good choice for the study of plasma chemistry. Mass spectrometry measures the plasma composition directly at the surface and is not limited by existence of accessible optical transitions. When properly designed and carefully calibrated mass spectrometry provides absolute densities of the measured species. It can even measure internally excited metastable species. Here, measurement of neutral species and positive ions generated in an atmospheric pressure plasmas jet operated with He, hexamethyldisiloxane and $O_2$ will be presented.

**Introduction**

Plasma processing with use of reactive gases is a powerful tool for material synthesis, removal or modification. The knowledge of absolute densities of species and their fluxes to the surface is crucial for understanding of plasma chemistry. Mass spectrometry is a diagnostic technique capable of measuring neutral species (including reactive radicals) and positive and negative ions [1-8]. The basics of mass spectrometry measurements with the focus on its application for the diagnostics of atmospheric pressure non equilibrium plasmas will be presented in the following.

**Mass spectrometry measurements**

Fig. 1 shows typical arrangements of the gas sampling from the plasma chamber. In the case of ion measurements, no ionization of the gas is necessary and therefore only ions originating directly from the plasma are measured. Differential pumping with only one pumping stage is sufficient as shown in Fig. 1(a). The signal is proportional to ion fluxes through the sampling orifice, not to ion densities directly. These fluxes are larger for lighter ions due to their higher velocity at the same ion density and ion energy. Additionally, other effects have to be considered in the quantitative analysis of the measured count rates. Positive ions are accelerated in the sheath in front of the reactor wall gaining up to few tens of electron volts kinetic energy. The acceptance angle of ions depends strongly on the ion energy and the ion optics of the MS may introduce chromatic aberration errors if the ion energy distribution function is not monoenergetic [9]. The transmission function of the MS itself, which discriminates heavier ions, has to be considered as well.

A more advanced arrangement is necessary for measurement of reactive neutral species. The MS signal $S_i$ of the species $i$ is directly proportional to particle density in the ionizer [1-3] and the background density of the species in the mass spectrometer chamber has to be taken into account. System with two or three pumping stages connected through aligned orifices is used in this case, which allows formation of an intense molecular beam (MB) and maintains low background pressure in the last stage with the MS, cf. Fig. 1(b). A beam-to-background ratio between the density in the beam and in the background gas of around 1 can be achieved in this way [1,3]. The background signal has still to be separated from the beam signal with the help of a mechanical chopper. The chopper can block the MB in front of the ionizer allowing measurement of the background signal only. This signal can then be subtracted from the signal measured without the chopper in blocking position. This arrangement allows absolute calibration of the densities of neutral species. It should be noted that the particle velocity does not play a role in measuring neutral species. A higher velocity (higher flux into the MS) is compensated by shorter residence time in the ionizer [2].

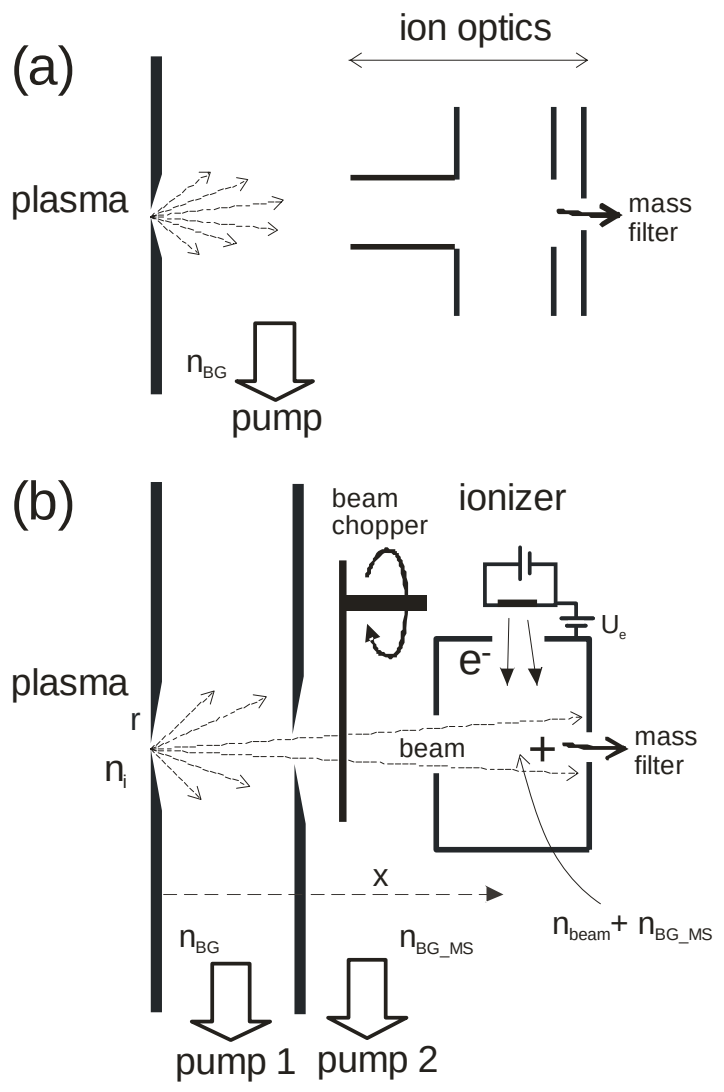

*Fig. 1 Arrangements of the gas sampling for detection of ions (a) and reactive neutral particles (b).*

The identification of the species detected at a given mass can be done by performing an electron energy scan at this mass in the energy region near the ionization energy (ionization threshold) of the detected species: threshold ionization mass spectrometry (TIMS). The electron impact ionization (EII) cross section is zero below this threshold energy, which is the ionization energy of the valence electron of a given species, and rises usually linearly just above this energy. The TIMS technique is especially important for detection of reactive dissociation products of a parent molecule, which originates directly from the plasma. It utilizes the difference of the EII threshold of a given radical and the electron impact dissociative ionization (DI) threshold of the parent molecule, which are typically several electron volts higher. Therefore, the radical can be detected by lowering the electron energy below the threshold for DI of the parent molecule. For example, detection of O in an $O_2$ discharge is possible with electron energies between 13.8 eV (direct ionization of O to $O^+$) and 18.0 eV (dissociative ionization of $O_2$ to $O^+$ + O) [4,10]. TIMS also allows detection of electronically excited species [5,11].

Mass spectrometry can also be used to analyze plasmas operated at atmospheric pressure [7,8,10,12]. Three pumping stages have to be usually used to reduce the pressure to an acceptable level for the operation of the MS. The big difference compared to analysis of low pressure plasmas is the fact that the gas sampling is collisional. The mean free path of species in low pressure plasmas (~ mm or more) is larger than the diameter of the sampling orifice (few hundreds μm) and the transport of the particles is in the molecular flow regime. Neither the plasma nor the concentrations of the particles in the molecular beam are disturbed. This is not the case at atmospheric pressure. The mean free path is below 100 nm and the sampling into the vacuum is collisional with the formation of a so called supersonic free jet [13]. The thermal energy of the particles is transformed in a directed movement by high pressure gradient at the sampling orifice and behind. Particles are accelerated to velocities higher than a sound velocity just behind the sampling orifice. The pressure in the first pumping stage is critical for the free jet expansion. Sampled particles in the free jet start to collide with the background gas if the pressure is too high and the mean free path is smaller than the length between the sampling orifice and the skimmer into the second stage; a so called barrel shock appears. The skimmer, which connects the first and second stage, has to be placed upstream from this barrel shock to minimize the change of the beam composition. When the background pressure in the first stage is good enough a transition of the free jet into a molecular beam occurs and the position of the skimmer is not critical.

Even under ideal conditions, the composition of the gas mixture in the molecular beam is changed in the collisional expansion compare to the composition in the plasma. This effect is called composition distortion in MBMS sampling and has been described by Knuth [14] in detail. One of the major effects is radial diffusion in the free jet depending on the mass of the particles and on the mean molecular weight of the gas mixture. A general trend can be recognized with lighter particles diffusing more easily out of the beam axis and hence being discriminated against heavier particles at the measuring point (ionizer). Additionally, a pressure gradient appears just in front of the sampling orifice, which can disturb the plasma.

**Sampling system with rotating skimmer**

The low pressure necessary for the undisturbed formation of the supersonic free jet can be realized by placing a special beam chopper into the first pumping stage [10,15]. It consists of a rotating stainless steel disk with small embedded skimmers as shown in Fig. 2. The penetration of the sampled particles into the second stage of the pumping system is blocked, if the embedded skimmer is not aligned with the other orifices. A good background pressure is generated in the second and third pumping stage in this phase because the gas conductivity "around" the rotating disk is very small. The gas enters directly the second and third stage only when the embedded skimmer is aligned with the sampling orifice. The supersonic free jet is formed in this case without disturbance from the background particles and an ideal molecular beam is formed for a short time. A high peak with a beam-to-background ratio of around 14 appears in the time resolved measurements of the detector current [10,15]. No additional beam chopper is necessary. Detection of neutral species in ppm concentrations is possible with this system. However, even if the supersonic free jet is expanding into the region of very good vacuum, a composition distortion inherent to the collisional sampling has to be considered.

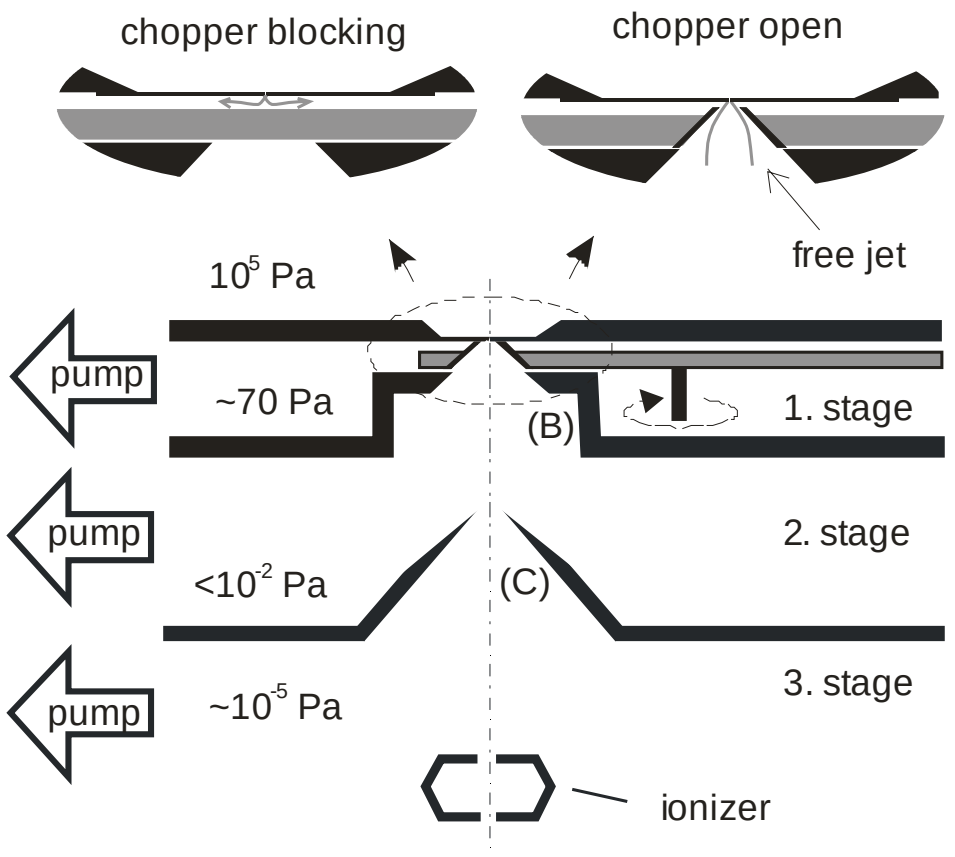

*Fig. 2. Scheme of the gas sampling system from the atmospheric pressure with the chopper with rotating skimmer in the first pumping stage.*

The sampling system introduced in Fig. 2 was used to analyze the composition of the effluent of a parallel plate microscale atmospheric pressure plasma jet (μ-APPJ) [10, 16]. This plasma source is formed by two parallel electrodes at 1 mm distance with 1 mm width and 30 mm length. The interelectrode region is confined by two glass plates on both sides forming a 30 mm long channel with a cross section of 1x1 $mm^2$. RF voltage of $U_{RMS}$ = 230 V is applied at 13.56 MHz frequency to the electrodes. The source is operated with 1 to 5 slm of He and up to 2 % of $O_2$. The absolute density of atomic oxygen measured by the MS is in very good agreement with values measured by two photon absorption laser induced fluorescence (TALIF) [10]. Additionally, the density of ozone ($O_3$) could be determined as well.

**Measurement of atmospheric pressure plasma with HMDSO molecules**

This system is now used to analyze the reaction pathways of hexamethyldisiloxane [HMDSO, $(CH_3)_3SiOSi(CH_3)_3$] in He plasma in the μ-APPJ source ($U_{RMS}$ = 230 V, electrode length 10 mm, jet-substrate distance 4 mm). The jet is placed in a small chamber with controlled He atmosphere, which is mounted on the front plate of the MS. HMDSO gas in mixtures with $O_2$ is commonly used as a precursor for deposition of $SiO_2$ films at low or atmospheric pressure [17-20]. We apply 0.1 sccm of HMDSO (20 ppm in the gas mixture) and 10 sccm of $O_2$ for deposition of $SiO_2$ like coatings by means of μ-APPJ. The addition of $O_2$ is necessary, carbon rich films are deposited without addition of $O_2$.

First, the analysis of neutral particles will be discussed. Fig. 3 shows the change of the MS signal at four masses when the plasma is switched on, measured with 20 eV electron energy in the ionizer to minimize dissociative ionization and fragmentation: 147 amu (highest peak in the fragmentation pattern of HMDSO), 75 amu (representing most probably trymethylsilanol), 133 amu (representing pentamethyldisiloxane), and 221 amu (representing octamethyltrisiloxan). It should be noted that parent ions of all of these compounds are unstable and release one $CH_3$ group upon ionization. Therefore, these masses are always 15 amu smaller than the molecular mass of corresponding molecule (e.g. 162–15 = 147 amu ion for HMDSO molecule). Measurements with plasma off and plasma on are always performed and gas mixture without and with 10 sccm of $O_2$ are compared. The measurement of the gas mixture without oxygen shows only a very weak effect on the precursor gas. The HMDSO consumption (mass 147 amu) is only around 6 % and the increase of the MS signal at the other masses is within the statistical errors of the measurements. The signals are very weak, because only 20 ppm of HMDSO is added into the gas mixture and only ~ 1 ppm of it is consumed. The stable products will therefore have densities close to the detection limit of our MS.

A significant increase in the HMDSO consumption to 16 % and higher signals at the other measured masses are observed when $O_2$ is added into the gas mixture. The formation of trymethylsilanol indicates the enhanced breakage of the Si-O bond of the HMDSO molecule. It is consistent with the increase of octamethyltrisiloxan (221 amu), which is a product of the reaction of $(CH_3)_3SiO$ fragment with a HMDSO molecule. It is not clear, whether for example atomic oxygen reacts at the Si-O bond, or the addition of $O_2$ changes the plasma parameters (plasma density, electron energy distribution function) in such a way that the electron impact dissociation of HMDSO becomes more effective. The increase of the pentamethyldisiloxane concentration (133 amu) shows, that Si-C bonds of the HMDSO molecules are attacked as well. A direct reaction of the methyl group with atomic oxygen or excited $O_2$ molecules seems the most probable process responsible for it. Even if the HMDSO consumption is increased by addition of $O_2$, there is no indication that some carbon free radicals, which would serve as growth precursors for the carbon free $SiO_2$-like coating, are produced in the gas phase. Indeed, the recent measurements with separate surface treatment by HMDSO plasma and $O_2$ plasma have shown that surface reactions are responsible for the carbon removal from the grown film [21].

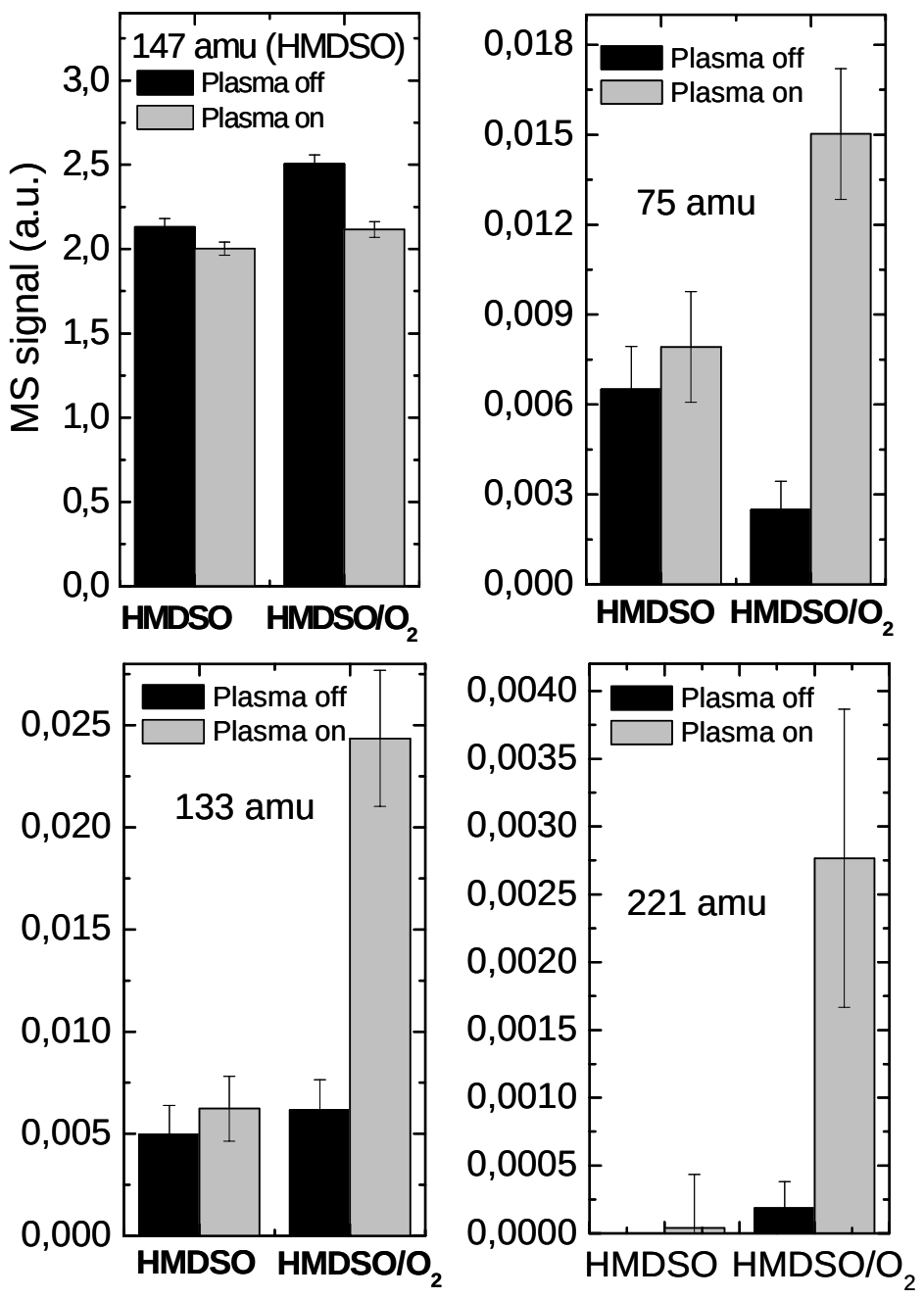

*Fig. 3 Neutral species detection, electron energy 20 eV. Change of MS signals at masses 75, 133, 147 and 221 amu under plasma off and plasma on conditions without and with $O_2$ flow. Jet-MS distance 4 mm, HMDSO flow 0.1 sccm, $O_2$ flow 10 sccm, He flow 5 slm.*

The MS system allows detection of positive ions as well. Since the active plasma is confined between electrodes and measurements are performed in the effluent, no ions are detected under conditions from Fig. 3 (0.1 sccm HMDSO, 4 mm jet-substrate distance). Many ions can however be detected at lower HMDSO flows. Fig. 4 shows the variation of the signal intensities as function of the jet-MS distance in a He/HMDSO plasma with 0.006 sccm of HMDSO. The ion spectrum is dominated by $H_3O^+(H_2O)_n$ clusters produced by sequential $H_2O$ molecule addition to a $H_3O^+$ ion. Their presence is typical for atmospheric pressure plasmas even under a controlled atmosphere [7,8]. The most abundant is the $H_3O^+(H_2O)_3$ cluster with a maximum at the distance of 5 mm from the jet, but still detectable at 19 mm. The fact that ions are generated in plasma effluent and quenched by increase of HMDSO flow indicates that they originate probably in secondary reactions of water or water clusters with either helium metastables or with plasma radiation. Additional measurements are planned to confirm this hypothesis. Two other ions have been observed, HMDSO·$H^+$ (163 amu) and HMDSO·$H_3O^+$ (181 amu). These ions are very probably formed in the plasma effluent in proton ($H^+$) transfer or ligand ($H_3O^+$) switching reactions of HMDSO molecules with $H_3O^+(H_2O)_n$ clusters [22].

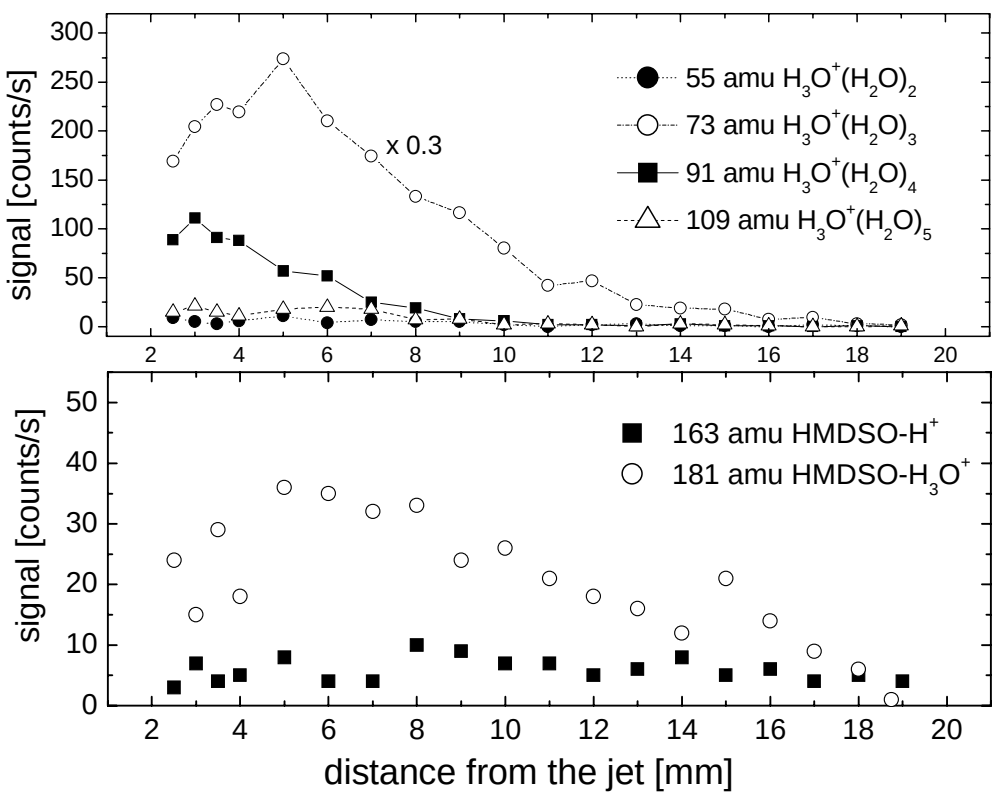

*Fig. 4 Measurement of positive ions as function of the distance from the jet. HMDSO flow 0.006 sccm.*

## Conclusions

Mass spectrometry is a powerful technique for measurement of the quantitative composition of reactive plasmas. When properly designed and calibrated it provides absolute densities of neutral stable and reactive species in the close vicinity of the surface and ion fluxes towards this surface. Moreover, the mass spectrometry can be successfully used for the analysis of atmospheric pressure plasmas. The measurement of neutral species and ions generated in the He/HMDSO(/$O_2$) plasma chemistry in a microscale atmospheric pressure plasma jet has been shown.

## Acknowledgement

The project is supported by the German science foundation in the research group FOR 1123, project C1.

## References

[1] H. Singh, J. Coburn, and D. Graves, "Mass spectrometric detection of reactive neutral species: Beam-to-background ratio," *J. Vac. Sci. Technol.*, 17, 2447, 1999
[2] H. Singh, J. Coburn, and D. Graves, "Appearance potential mass spectrometry: Discrimination of dissociative ionization products," *J. Vac. Sci. Technol.*, 18, 299, 2000
[3] J. Benedikt, S. Agarwal, D. J. Eijkman, W. Vandamme, M. Creatore, and M. C. M. van de Sanden, "Threshold ionization mass spectrometry study of hydrogenated amorphous carbon films growth precursors," *J. Vac. Sci. Technol.* A, 23, 1400, 2005
[4] S. Agarwal, G. W. W. Quax, M. C. M. van de Sanden, D. Maroudas, and E. S. Aydil, "Measurement of absolute radical densities in a plasma using modulated-beam line-of-sight threshold ionization mass spectrometry," *J. Vac. Sci. Technol.* A, 22, 71, 2004
[5] S. Agarwal, B. Hoex, M. C. M. van de Sanden, D. Maroudas, , and E. S. Aydil, "Absolute densities of N and excited $N_2$ in a $N_2$ plasma," *Appl. Phys. Lett.*, 83, 4918, 2003
[6] J. Benedikt, D. C. Schram, and M. C. M. van de Sanden, "Detailed TIMS Study of Ar/C2H2 Expanding Thermal Plasma: Identification of a-C:H Film Growth Precursors," *J. Phys. Chem. A*, 109, 10153, 2005
[7] J. D. Skalny, J. Orszagh, N.J. Mason, J. Rees, Y. Aranda-Gonzalvo, T.D. Whitmore, "A mass spectrometric study of ions extracted from a point-to-plane dc corona discharge in N2O at atmospheric pressure," *J. Phys. D, 41*, 085202, 2008
[8] J.D. Skalny, J. Orszagh, N.J. Mason, J. Rees, Y. Aranda-Gonzalvo, T.D. Whitmore, "Mass spectrometric study of negative ions extracted from point to plane negative corona discharge in ambient air at atmospheric pressure," *Int. J. Mass Spectrom*., 272, 12, 2008
[9] E. Hamers, W. van Sark J. Bezemer, W. Goedheer, W. van der Weg, "On the transmission function of an ion-energy and mass spectrometer," *Int. J. Mass. Spectrom*., 173, 91, 1998
[10] D. Ellerweg, J. Benedikt, A. von Keudell, N. Knacke, and V. Schulz von der Gathen, "Characterization of the effluent of a He/$O_2$ microscale atmospheric pressure plasma jet by quantitative molecular beam mass spectrometry," *New J. Phys*., 12, 013021, 2010
[11] J. Benedikt, Ch. Flötgen, A. von Keudell, "Threshold Ionisation Mass Spectrometry detection of reactive species in plasmas," *53rd Annual Technical Conference Proceedings of the Society of Vacuum Coaters*, pp. , 2010
[12] E. Stoffels, Y. Aranda-Gonzalvo, T.D. Whitmore, D.L. Seymour, J. Rees, „Mass spectrometric detection of short-living radicals produced by a plasma needle," *Plasma Sour. Sci. Technol., 16*(3), 549, 2007
[13] "Atomic and Molecular Beam Methods," vol I, G. Scoles, Oxford: Oxford University Press, 1988
[14] E.L. Knuth, "Composition Distortion in MBMS Sampling," *Combust. Flame,* 103, 171, 1995
[15] J. Benedikt, D. Ellerweg, A. von Keudell, "Molecular beam sampling system with very high beam-to-background ratio: the rotating skimmer concept," *Rev. Sci. Instrum.,* 80, 055107, 2009
[16] N. Knake, S. Reuter, K. Niemi, V. Schulz-von der Gathen, J. Winter, Absolute atomic oxygen density distributions in the effluent of a microscale atmospheric pressure plasma jet. *J Phys. D*, 41, 194006, 2008

[17] V. Raballand, J. Benedikt, S. Hofmann, M. Zimmermann, A. von Keudell, „Deposition of silicon dioxide films using an atmospheric pressure microplasma jet,“ *J. Appl. Phys.*, 105, 083304, 2009
[18] D. Hegemann, H. Brunner, C. Oehr, „Evaluation of deposition conditions to design plasma coatings like $SiO_x$ and a-C:H on polymers,“ *Surf. Coat. Technol., 175*, 253, 2003
[19] M. Creatore, F. Palumbo, R. d'Agostino, P. Fayet, „RF plasma deposition of $SiO_2$-like films: plasma phase diagnostics and gas barrier film properties optimisation,“ *Surf. Coat. Technol.*, 142-144, 163, 2002
[20] S.E. Alexandrov, N. McSporran, M.L. Hitchman, „Remote AP-PECVD of Silicon Dioxide Films from Hexamethyldisiloxane (HMDSO),“ *Chem. Vapor Depos.*, 11, 481, 2005
[21] R. Reuter, D. Ellerweg, A. von Keudell, J. Benedikt, Surface reactions as carbon removal mechanism in deposition of silicon dioxide films at atmospheric pressure, Accepted for publication in *Appl. Phys. Letters* (2011)
[22] Blake, R. S., Monks, P. S., Ellis, A. M. Proton-transfer reaction mass spectrometry. *Chemical reviews*, 109, 861, 2009